\documentclass[
12pt,
aps,
a4paper,
amsmath,
superscriptaddress,
showpacs,
floatfix
]{revtex4-1}

\newif\ifOnecolumn
 \ifx\Onecolumn\undefined
 \Onecolumntrue
\else
 \Onecolumnfalse
\fi

\usepackage{graphicx,amssymb,amsmath,amsfonts}
\usepackage[mathscr]{eucal}
\usepackage{bm}

\usepackage{comment}
\usepackage{verbatim}

\let \ShowFixme = 1
\let \Date = 1

\newcommand{\rr}{\mathbf{r}}

\begin{document}

\title{Superionic state in double-layer capacitors with nanoporous electrodes}

\author{S.~Kondrat}
\email[Electronic address: ]{skondrat@ic.ac.uk}
\affiliation{Max-Planck-Institut f\"ur Mathematik in den Naturwissenschaften, D-04103 Leipzig, Germany}
\affiliation{Department of Chemistry, Faculty of Natural Sciences, Imperial College London, SW7 2AZ, UK}

\author{A.~Kornyshev}
\email[Electronic address: ]{a.kornyshev@imperial.ac.uk}
\affiliation{Department of Chemistry, Faculty of Natural Sciences, Imperial College London, SW7 2AZ, UK}
\begin{abstract}

In the recent experiments [Chmiola \textit{et al}, Science \textbf{313}, 1760 (2006); Largeot \textit{et al}, J. Am. Chem. Soc. \textbf{130}, 2730 (2008)] an anomalous increase of the capacitance with a decrease of the pore size of a carbon-based porous electric double-layer capacitor has been observed. We explain this effect by the image forces which exponentially screen out the electrostatic interactions of ions in the interior of a pore. Packing of ions of the same sign becomes easier and is mainly limited by steric interactions. We call this state `superionic' and suggest a simple model to describe it. The model reveals a possibility of a voltage-induced first-order transition between a cation(anion)-deficient phase and a cation(anion)-rich phase which manifests itself in a jump of capacitance as a function of voltage. 


\end{abstract}

\date{\today}

\maketitle

Electric double layer supercapacitors (EDLCs) have recently attracted a considerable attention due to their high energy density and fast power delivery \cite{conway:99,*arico:05,*woolfe:05, gogotsi:phil:10}. The carbon-based EDLCs store energy at the electrolyte/carbon interface, and in order to increase the amount of the energy stored it is necessary to increase the carbon surface area. This is usually achieved by using highly porosous carbon materials, with the pore size being \emph{twice} the size of the ions to allow ions adsorption on both pore walls~\cite{frackowiak:07, gogotsi:phil:10}.

In the recent experiments \cite{gogotsi:sci:06, gogotsi:08} Gogotsi and co-workers have measured the capacitance per surface area of EDLCs with nanoporous carbon-based electrodes \emph{vs} the pore width. They obtained a surprising result: \emph{an anomalous increase of the capacitance with decreasing the pore width} for nano and subnano pores comparable in size with the size of bare ions. Both an organic electrolyte~\cite{gogotsi:sci:06} and a (solvent-free) ionic liquid~\cite{gogotsi:08} have been tested as electrolyte medium leading to the same effect.

Molecular Dynamics (MD) simulations have confirmed the experimental results \cite{yang:09, shim:10}. For instance, in Ref.~\cite{shim:10} the simulations have been performed for EMI cations and TSFI anions as ionic liquid and the interior of a single-wall carbon nanotube (CNT) as pore. The `anomalous' increase of the capacitance has been observed for the CNT radii down to the ions size.
   
It has been proposed that partial desolvation of the ions entering small pores is responsible for the anomalous behaviour of the capacitance in the case of organic electrolytes~\cite{gogotsi:sci:06}. A recently considered electric wire-in-cylinder model \cite{meunier:07, *meunier:08} offers a good fit to the experimental data but for further development of the porous EDLCs it is important to understand the underlying physics of the capacitance in nanopores.

In this Letter we propose a simple explanation of the anomalous capacitance behaviour by introducing the concept of a \emph{superionic state in a metallic nanopore} caused by (i) exponential screening of the ion-ion pair interactions in a nanogap, which arises due to the image forces and enables more ions of the same sort to occupy the pore. By the same token, there is also (ii) an electrostatic contribution to the free energy of transfer of ions from the bulk to the pore interior. The latter tends to increase the ion density in a pore, compensating the opposing trend due to the loss of solvation shell by the ions moving into the pore.

\begin{figure}[t]
    \begin{center}
	\ifOnecolumn
        	\includegraphics*[width=0.25\textwidth]{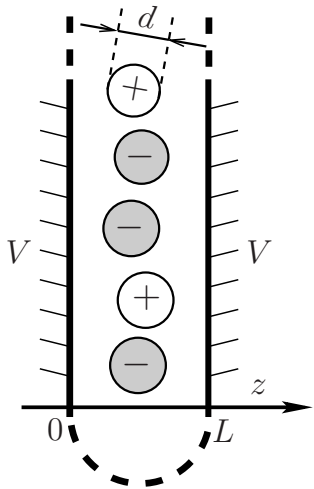}
	\else
	 	\includegraphics*[width=0.15\textwidth]{model}
	\fi
        \caption{
        \label{fig:model}
        Schematic view of the cross section of a single, laterally infinite, slit-like narrow pore as a part of a porous electrode. The pore width is $L$, $V$ is the voltage with respect to the reference electrode (in the ``bulk'' outside the pore), and $d$ is the diameter of ions.}
\end{center}
\end{figure}

Let us consider a point charge at $z=z_1$ confined between two metal plates (c.f. Fig.~\ref{fig:model} with $V=0$). The electrostatic potential satisfying the Laplace equation and Dirichlet boundary conditions at the plates can be easily found using the Fourier--Bessel transform to give
\ifOnecolumn
\begin{align}
\label{eq:pot}
\phi (z, R;z_1)
   = \frac{4}{\varepsilon L} \sum_{n=1}^\infty \sin(\pi n z_1/L) \sin(\pi n z/L) \;
     K_0(\pi n R/L),
\end{align}
\else
\begin{multline}
\label{eq:pot}
\phi (z, R;z_1)
   = \frac{4}{\varepsilon L} \sum_{n=1}^\infty \sin(\pi n z_1/L) \; \sin(\pi n z/L) \\
    \times K_0(\pi n R/L),
\end{multline}
\fi
where $\varepsilon$ is the dielectric constant of a medium between the plates, $R$ is the lateral distance from the charge, $L$ is the plates separation (i.e., the pore size), and $K_n(x)$ is the modified Bessel function of the second kind of order $n$. Using the asymptotic behaviour of $K_0(x)$ for $x \gg 1$~\cite{gradstein:81} one finds for $R \gg L/\pi$
\begin{align}
    \label{eq:pot:exp}
    \phi (z, R;z_1) \approx  \frac{e^{-\pi R/L}}{\varepsilon\sqrt{ 2LR }} \; \sin(\pi z_1/L) \sin(\pi z/L),
\end{align}
which means that the electric potential exponentially decays with the distance in the plane parallel to the plates, and the Coulomb interaction between two charges is effectively \emph{screened out}. In the opposite limit of large $L \gg \pi R$, Eq.~(\ref{eq:pot}) recovers the  Coulomb law.

We consider for simplicity a single, metallic~\cite{*[{In the experiments of Ref.~\cite{gogotsi:sci:06, gogotsi:08} the carbide-derived carbon (CDC) materials have been used for electrodes. The CDCs are characterized by the properties close to the ones of metals, in particular by relatively high conductivity. In our work we use the simplest approximation, assuming that the pore walls are perfect conductors so that the electric field vanishes inside the walls. For semi-metals, like graphite, the electric field penetrates the walls and decays exponentially there, see }] [{. In the present context it means that the wall separation will be, loosely speaking, effectively shifted by some small $\delta L>0$, which shall slightly weaken the effects discussed here.}] kornyshev:77}, slit-like~\footnote{Cylindrical pores have been used in the experiments and simulations~\cite{gogotsi:sci:06, gogotsi:08, yang:09, shim:10}. In a cylindrical pore the effect of screening is even stronger than in a slit pore. A slit pore, in turn, is a first approximation for connected-pores geometries (such as the space between nanotubes), and can also be beneficial for fast charging/discharging dynamics.} pore, infinitely extended in the lateral directions (see Fig.~\ref{fig:model}), and neglect the boundary effects at the  ``entrance'' and ``exit'' or closing of the pore (for supercapacitor electrodes the typical pore length is usually much longer then its width). The electrostatic potential at the pore walls is measured with respect to the reference electrode in the `bulk' of the capacitor. We consider a small nanoscale pore, in which there is room only for one ionic layer (as in the experiments and simulations). Furthermore, we assume for simplicity that the ions are situated on average in the middle of the pore (i.e., $z=L/2$, c.f.~Eqs.~(\ref{eq:U}) and (\ref{eq:self})). Although not generally true, this assumption should not qualitatively change our conclusions.

Using now Eq.~(\ref{eq:pot}), and taking the voltage into consideration, one finds for the internal energy (per surface area) due to electrostatic interactions of an ionic liquid in a pore
\ifOnecolumn
\begin{align}
    \label{eq:U}
    \beta U (\rho, c) = \beta e V c + 4 c^2 \;R_c(\rho) L_B \sum_{m=1}^\infty \frac{\sin^2 (\pi m /2)}{m^2}
    K_1 \bm{(}\pi m R_c(\rho)/L\bm{)},
\end{align}
\else
\begin{multline}
    \label{eq:U}
    \beta U (\rho, c) = \beta e V c +  
    4 c^2 \;R_c(\rho) L_B \sum_{m=1}^\infty \frac{\sin^2 (\pi m /2)}{m^2}\\
    \times K_1 \bm{(}\pi m R_c(\rho)/L\bm{)},
\end{multline}
\fi
where $\beta= 1/k_B T$, with $k_B$ being the Boltzmann constant and $T$ the temperature, and $L_B=\beta e^2/\varepsilon$ is the Bjerrum length~\footnote{\label{endnote:bjerrum}The Bjerrum length is the distance between two charges at which their Coulomb interaction energy in a medium of a given dielectric constant $\varepsilon$ is equal to the thermal energy. The value of the Bjerrum length in the bulk is of order of $5$~nm for a typical ionic liquid and $0.7$~nm for an aqueous electrolyte solution at room temperature. In nano-pores the dielectric constant is reduced, being determined by an effective polarizability of a quasi two-dimensional layer of ions. Its value is not known but we expect it to be in the range between $2$ and $5$, which gives the Bjerrum length between $11$ and $28$~nm.}; $c = Z_+ \rho_+ - Z_- \rho_-$ and $\rho = \rho_+ + \rho_-$ are the \emph{two-dimensional} charge density and total density of the ions in the pore, respectively, where $\rho_\pm$ are the (two-dimensional) densities of the $\pm$ ions and $Z_\pm$ the corresponding valencies. In the derivation of Eq.~(\ref{eq:U}) we have used the cut-out disk approximation~\cite{kornyshev:jeac:86} with the cut-out radius $R_c(\rho)=(\pi \rho)^{-1/2}$. The voltage, which appears linearly in Eq.~(\ref{eq:U}), drives the anions to and cations from the pore for $V>0$, and vice versa for $V<0$.

We present the free energy of transfer of an ion from the bulk into a pore as a sum of two contributions: (i) a ``re-solvation'' free energy $\delta E_{\alpha}$ ($\alpha=\{+,-\}$) due to the (partial) loss of the solvation shell when an ion moves from the bulk to the pore, which we assume to be $L$-independent, and (ii) a change in the self-energy of an ion due to confinement, which characterizes its interaction with the metal plates. The latter can be approximated as the work of ``charging'' of a \emph{point charge} inside the pore minus the work needed to charge it in the bulk with the same dielectric constant: $(q^2/2)\;\lim_{\rr \to \rr_1} \bm{(} \phi_L(\rr, \rr_1) - \phi_{L=\infty}(\rr,\rr_1)\bm{)}$. Using Eq.~(\ref{eq:pot}) one obtains for this contribution to the internal energy (per surface area)
\begin{align}
\label{eq:self}
\beta E_s(\rho_\pm) =  \sum_{\alpha=\pm} \Big( \beta \delta E_\alpha 
	- L_B/L\,f(1/2)\, Z_\alpha^2 \Big) \rho_\alpha,
\end{align}
where
\begin{align}
    f(x) = \int_0^\infty
    \left(\frac{1}{2}  - \frac{\sinh\bm{(}Q(1-x)\bm{)}\sinh(Qx)}{\sinh(Q)}\right) d Q.
\end{align}
The function $f(x)$ is positive-definite around $f(x=1/2) = \min_{x \in[0;1]} f(x) =\ln(2)$, and hence the second term in Eq.~(\ref{eq:self}) is negative. The resolvation energy $\delta E_\pm$ is positive and depends on the type of electrolyte medium; its value is not known exactly but we estimate it to be not larger than $20\,k_B T$~\footnote{Such estimate can be obtained by combining the methods of calculation of solvation energy in condensed media~\cite{kornyshev:85} and typical forms for nonlocal dielectric function of ionic liquids that accounts for overscreening~\cite{tossi}}.This implies that for sufficiently large $L_B/L$ the total free energy of transfer is negative, and will therefore favour an increase of the ion densities inside the pore relative to the bulk (irrespective of the voltage and the sign of the charges).

The total free energy of an ionic liquid in the pore (per surface area) is
\begin{align}
    F = U + E_s + \sum_{\alpha=\pm} \mu_\alpha \rho_\alpha - T S,
    \label{eq:fe}
\end{align}
where $S$ is the entropy and $\mu_\pm$ chemical potentials. For the ions of the same size $d$ the entropy is given by~\cite{borukhov:97, *kornyshev:07}
\ifOnecolumn
\begin{align}
    \label{eq:entropy}
    S = -k_B \left\{
    \sum_{\alpha=\pm} \rho_\alpha \ln \left( \frac{\rho_\alpha v_0}{L} \right)
        +  \frac{L}{v_0} \left(1 - \frac{\rho v_0}{L}\right)
        \ln \left(1 - \frac{\rho v_0}{L} \right)
    \right\},
\end{align}
\else
\begin{multline}
    \label{eq:entropy}
    S = -k_B \left\{
    \sum_{\alpha=\pm} \rho_\alpha \ln \left( \frac{\rho_\alpha v_0}{L} \right)
        \right. \\
    + \left. \frac{L}{v_0} \left(1 - \frac{\rho v_0}{L}\right)
        \ln \left(1 - \frac{\rho v_0}{L} \right)
    \right\},
\end{multline}
\fi
where $v_0 = \pi d^3/6 \eta_m$ is the minimal volume per ion and $\eta_m$ the maximum packing fraction~\cite{*[{The maximum packing fraction $\eta_m=\pi/6$ for a close-packed simple cubic lattice and $\sqrt{2}\pi/6$ for a face-centered cubic or hexagonal lattice, and $\eta_m\approx 0.64$ for random close-packing, see e.g. }] [{. We use $\eta_m=\pi/6$ here and note that other values of $\eta_m$ will merely shift the capacitance curves and locations of the transitions discussed here but shall not lead to any qualitative changes.}] song:08}. The first term in Eq.~(\ref{eq:entropy}) corresponds to the entropy of ions, and the last term is the entropy of a solvent (if there is one). Note that Eq.~(\ref{eq:entropy}) reflects the three-dimensionality of the problem~\footnote{We allow the total density to exceed the maximum two-dimensional density and limit its value by $L/ v_0$ instead. In this way we mimic possible deviations of the ion locations from the middle of the pore. The effects discussed below do persist if we use the fully two-dimensional entropy but the values obtained are of course different.}. Finally, the chemical potentials $\mu_\pm$ are set to the chemical potentials of the ionic liquid in the ``bulk'', i.e., where the electric potential vanishes (we recall that, besides here, the bulk part does not appear in the model); one gets
\begin{align}
    -\beta \mu_\pm  = \ln\left( \frac{v_0 \bar \rho_\pm^{(\infty)}}{1- v_0 \bar \rho_\infty} \right)
    = \ln\left(\frac{\gamma \; Z_\pm }{Z(1-\gamma)}\right),
\end{align}
where $Z=Z_+ + Z_-$, $\bar \rho_\infty = \bar \rho_+^{(\infty)} + \bar \rho_-^{(\infty)}$ is the total (three-dimensional) density of the ions in the bulk, $\gamma = v_0 \bar \rho_\infty$ (see Ref.~[\onlinecite{borukhov:97, *kornyshev:07}]), and we have used the electro-neutrality condition obeyed in the bulk.

The free energy (\ref{eq:fe}) can now be minimized numerically with respect to the ion densities $\rho_\pm$, and the differential capacitance can be found by numerically differentiating the negative of the total charge accumulated in the pore with respect to the voltage.

\begin{figure}[t]
    \begin{center}
	\ifOnecolumn
        	\includegraphics*[width=0.55\textwidth]{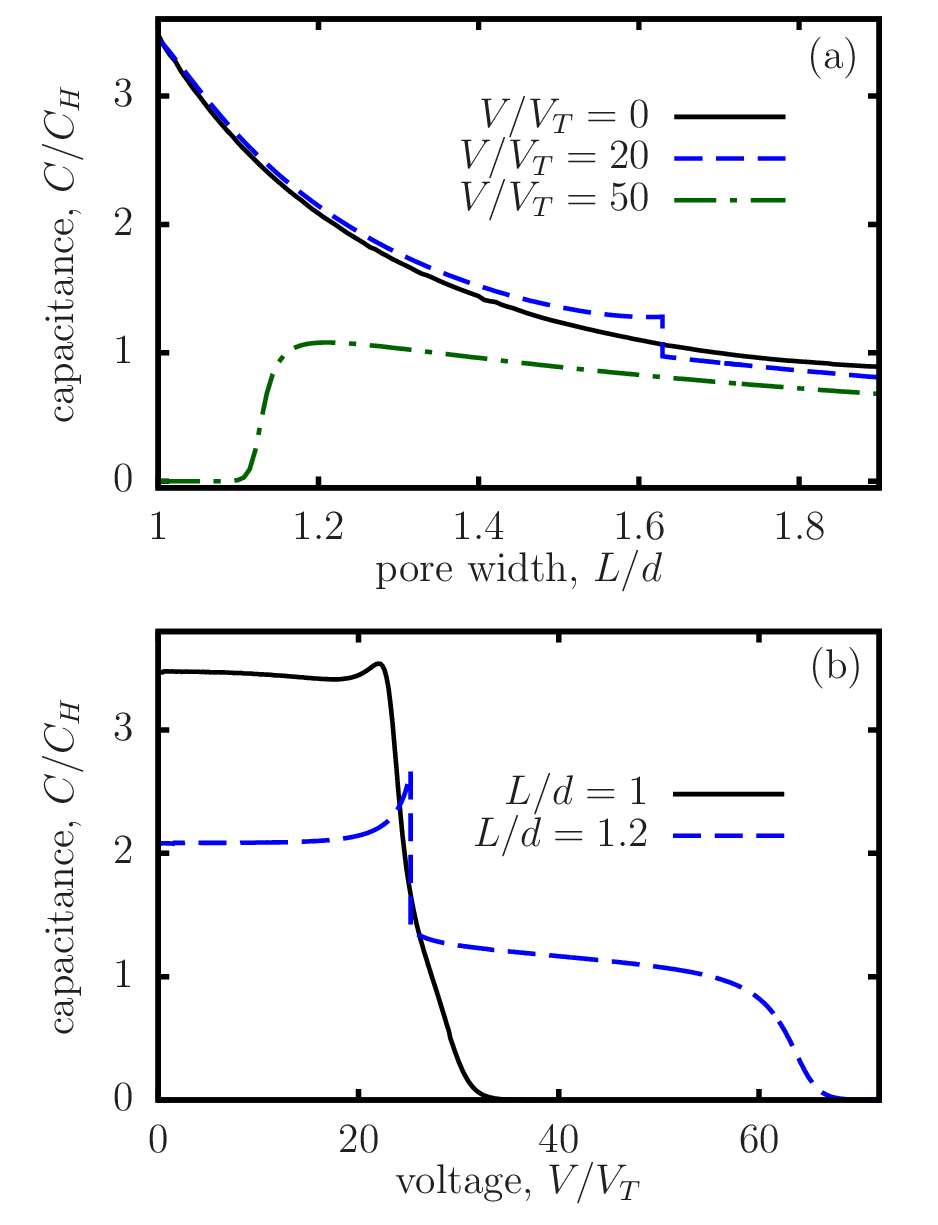}
	\else
	        \includegraphics*[width=0.34\textwidth]{Cap}
	\fi
        \caption{
        \label{fig:cap}
        (color online) Differential capacitance per surface area, expressed in terms of the Helmholtz capacitance $C_H=\varepsilon/2\pi d$, as a function of (a) pore width $L/d$ and (b) voltage $V/V_T$, where $V_T = 1/ \beta e$ is the thermal voltage ($\approx 26$~mV at room temperature). The ion diameter is $d=0.7$~nm, Bjerrum length $L_B=20$~nm, $\eta_m = \pi/6$, $\gamma=0.5$, re-solvation energies $\delta E_\pm = 10\,k_BT$, and valencies $Z_\pm =1$. The capacitance increases with decreasing the pore width at zero voltage (solid line in (a)), in accordance with the experimental observations (see Refs.~\cite{gogotsi:sci:06, gogotsi:08}). There is a jump in capacitance at $L/d \approx 1.63$ for $V/V_T=20$ in (a) and at $V/V_T \approx 25.2$ for $L/d=1.2$ in (b) manifesting a first order transition. A maximum for $L/d=1$ in (b) reflects a close vicinity of a critical end-point (c.f.~Fig.~\ref{fig:pd}).}
    \end{center}
\end{figure}

The capacitance (per surface area) \emph {vs} the pore width is shown in Fig.~\ref{fig:cap}(a) for a few values of the voltage $V$. At zero voltage the capacitance \emph{increases with decreasing the pore width} (solid line in Fig.~\ref{fig:cap}(a)), in agreement with the experimental observations~\cite{gogotsi:sci:06, gogotsi:08}. Note that $C > C_H$  for small pore widths, where $C_H=\varepsilon/2\pi d$ is the Helmholtz capacitance~\footnote{A similar `enhancement' of the double layer capacitance has recently been reported for glassy electrolytes and planar electrodes~\cite{mariappan:10}, and has been claimed to result from the reduced ion-ion interactions at the surface~\cite{shklovskii:10:prl}. We note that in nanogaps such a reduction is \emph{exponential} and has different manifestations.}. Obviously, $C\equiv 0$ for pores smaller than the ion size because the ions cannot enter the pore (not shown in Fig.~\ref{fig:cap}). The differential capacitance also vanishes for voltages higher than a certain threshold voltage $V_0(L)$ (see Fig.~\ref{fig:cap}(b) and dot-dash line in Fig.~\ref{fig:cap}(a)); this is because the pore is fully occupied by anions for $V>V_0$.

\begin{figure}[t]
    \begin{center}
	\ifOnecolumn
	        \includegraphics*[width=0.55\textwidth]{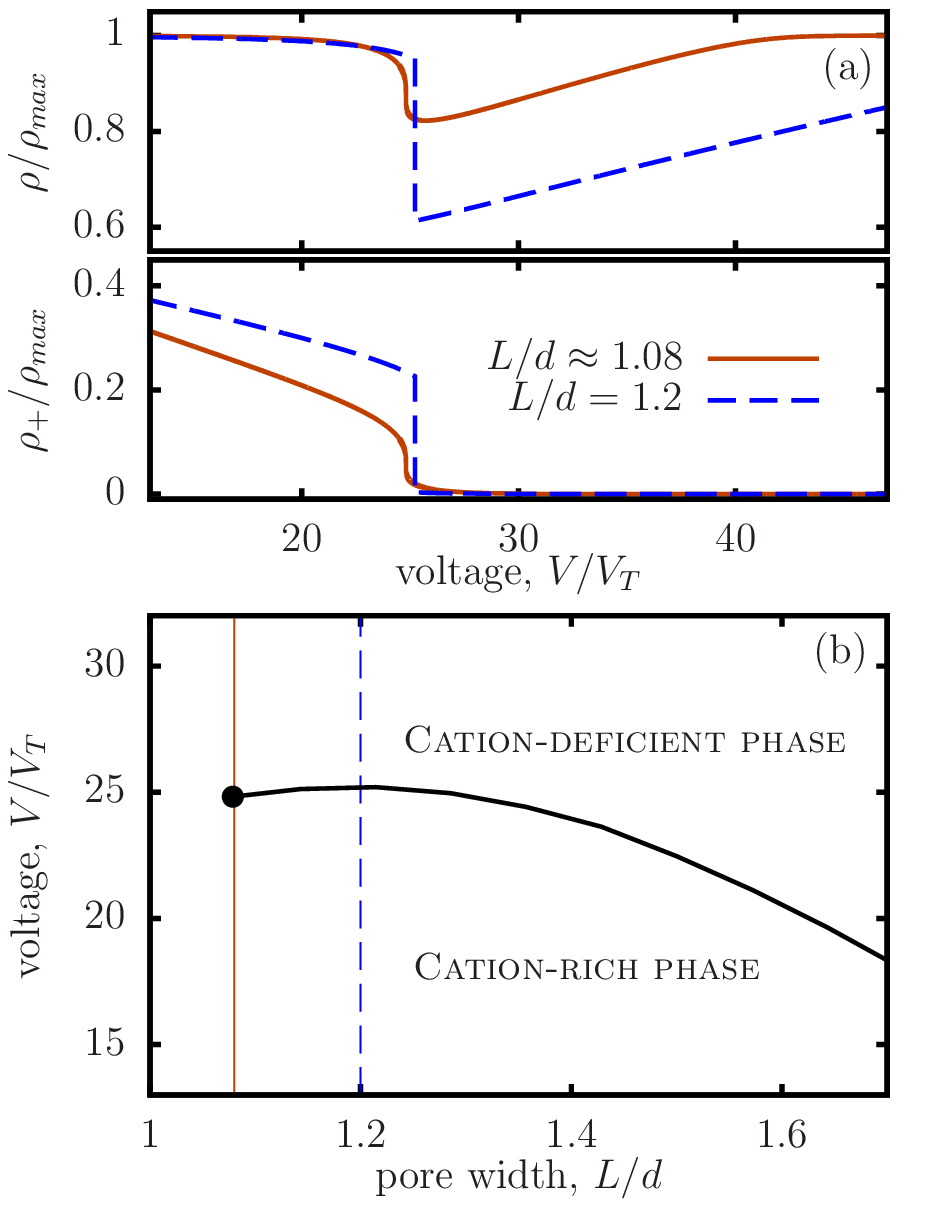}
	\else
        	\includegraphics*[width=0.34\textwidth]{PD}
	\fi
        \caption{
        \label{fig:pd}
        (color online) (a) The total density $\rho$ and the cation density $\rho_+$, expressed in terms of $\rho_{max} = L/v_0$, as a function of voltage for the pore width corresponding to a first order transition ($L/d=1.2$) and a critical end-point ($L/d \approx 1.08$). (b)~The phase diagram spanned in the (voltage, pore size) plane. The solid line corresponds to a first order transition between a cation-deficient phase and a denser, cation-rich phase, and the full circle denotes a critical end-point. Thin vertical lines correspond to the profiles in (a). The model parameters are the same as in Fig.~\ref{fig:cap}.}
    \end{center}
\end{figure}

For larger pores and intermediate voltages we observe a \emph{jump} in both capacitance \textit{vs} pore size and capacitance \textit{vs} voltage curves (dash lines in Fig.~\ref{fig:cap}). It marks the onset of a \emph{first order phase transition}. This transition occurs when the applied voltage (positive in this case) is sufficiently high to drive the cations out of the pore but is not yet high enough to fill the pore completely with anions. In other words, the transition separates a cation-deficient (CD) phase and a denser cation-rich (CR) phase. (In the case $V<0$ the transition separates the anion-rich (AR) and anion-deficient (AD) phases.) This is demonstrated in Fig.~\ref{fig:pd}(a) where we plot the cation density $\rho_+$ \emph{vs} voltage. At $L/d=1.2$ the density profile $\rho_+(V)$ is discontinuous, manifesting a first order transition. At smaller $L/d\approx 1.08$ the profile is continuous but $d\rho_+ /dV$ diverges at $V/V_T \approx 24.8$ which indicates a critical end-point. The corresponding phase diagram in the (voltage, pore-size) plane is shown in Fig.~\ref{fig:pd}(b).
 
The CR/AR phases are characterized by high total ion density close to $\rho_{max} = L/v_0$ (see Fig.~\ref{fig:pd}(a)). This is due to the superionic state created in the pore interior, i.e., the favourable free energy of transfer [see Eq.~(\ref{eq:self})] and the exponential screening of the Coulomb interactions [see Eqs.~(\ref{eq:pot:exp}) and (\ref{eq:U})]. The pore charging in this case is mainly due to the \emph{exchange} of the cations in the pore with the anions from the outside of the pore (for $V>0$) such that the total density remains practically constant (see Fig.~\ref{fig:pd}(a)). This is to be contrasted with the charging in the CD and AD phases, which is essentially characterized by an increase of the total ion density. Therefore, the phase transition lines also separate two different `charging regimes.' It is reasonable to expect the charging/discharging dynamics to be also different in these two regimes, which may be of a considerable importance for power delivery of porous EDLCs.

In summary, we have considered a simple phenomenological model in order to understand the properties of EDLCs with nanoporous electrodes. Our model takes into account the exponential screening of the ion-ion interactions in a metallic nanopore and the interaction of ions with pore walls -- both determined by image forces, which underpin what we call a `superionic state' in metallic nanopores. It explains an `anomalous' increase of the capacitance with a decrease of the pore size~\cite{gogotsi:sci:06, gogotsi:08}. We have also predicted an interesting voltage-induced first-order transition between a cation(anion)-deficient phase and a denser cation(anion)-rich phase, manifested as a jump in the capacitance. In spite of the coarse nature of the model and a number of simplifying assumptions, we believe it leads to qualitatively correct results. It would be beneficial to check its predictions using, for instance, a more robust density-functional theory or Monte Carlo/Molecular Dynamics simulations, which can also probe the discussed effects in a wider range of pore sizes, and of course to verify them by experiments.

The authors thank Prof.~Yury Gogotsi and Dr.~Maxim Fedorov for useful discussions. Support by the Grant  EP/H004319/1 of EPSRC is gratefully acknowledged.

%\bibliographystyle{apsrev}
%\bibliography{manuscript}


\end{document}